\documentclass[fleqn,usenatbib]{mnras}

\usepackage[T1]{fontenc}
\usepackage{ae,aecompl}

\usepackage{graphicx}	
\usepackage{amsmath}	
\usepackage{amssymb}	

\newcommand{\kms}{\hbox{km s$^{-1}$}}
\newcommand{\re}{\hbox{${\rm R}_{\rm e}$}}

\newcommand{\msun}{\hbox{M$_{\odot}$}}
\newcommand{\dgs}{DGSAT~I}

\title[Chemical abundance ratio of an UDG]{Extreme chemical abundance ratio suggesting an exotic origin for an ultra-diffuse galaxy}

\author[I. Mart\'in-Navarro]{
Ignacio Mart\'in-Navarro$^{1,2}$\thanks{E-mail: imartinn@ucsc.edu}, Aaron J. Romanowsky$^{1,3}$, Jean P. Brodie$^{1}$ \newauthor
Anna Ferr\'e-Mateu$^{4,5}$, Adebusola Alabi$^{1}$, Duncan A. Forbes$^{4}$, Margarita Sharina$^{6}$ \newauthor
Alexa Villaume$^{1}$, Viraj Pandya$^{1}$, and David Martinez-Delgado$^{7}$
\\
$^{1}$University of California Observatories, 1156 High Street, Santa Cruz, CA 95064, USA\\
$^{2}$Max-Planck Institut f\"ur Astronomie, Konigstuhl 17, D-69117 Heidelberg, Germany\\
$^{3}$Department of Physics \& Astronomy, San Jos\'e State University, One Washington Square, San Jose, CA 95192, USA\\
$^{4}$Centre for Astrophysics \& Supercomputing, Swinburne University of Technology, Hawthorn VIC 3122, Australia\\
$^{5}$Institut de Ciencies del Cosmos (ICCUB),Universitat de Barcelona (IEEC-UB), E02028 Barcelona, Spain\\
$^{6}$Special Astrophysical Observatory, Russian Academy of Sciences, N. Arkhyz, KChR 369167, Russia\\
$^{7}$Astronomisches Rechen-Institut, Zentrum f{\"u}r Astronomie, Universit{\"a}t Heidelberg, M{\"o}nchhofstr. 12--14, 69120\\
}

\date{Accepted XXX. Received YYY; in original form ZZZ}

\pubyear{2018}

\begin{document}
\label{firstpage}
\pagerange{\pageref{firstpage}--\pageref{lastpage}}
\maketitle

\begin{abstract}

Ultra diffuse galaxies are a population of extended galaxies but with relatively low luminosities. The origin of these objects remains unclear, largely due to the observational challenges of the low surface brightness Universe. We present here a detailed stellar population analysis of a relatively-isolated UDG, \dgs, based on spectroscopic data from the Keck Cosmic Web Imager integral field unit. The star formation history of \dgs \ seems to be extended, with a mean luminosity-weighted age of $\sim$3 Gyr, in agreement with previous photometric studies. However, we find a very high [Mg/Fe] abundance ratio, which is extreme even in the context of the highly alpha-enhanced massive ellipticals and ultra-faint dwarfs. The [Mg/Fe]-enhancement of \dgs \ appear to be 10 times higher than the most magnesium-enhanced stellar systems discovered to date, and suggests that the chemical enrichment of this object was dominated by core-collapse supernovae. Intriguingly, this breaks the canonical relation between [Mg/Fe] and star formation time-scale. With a measured velocity dispersion of $56 \pm 10$ \kms, \dgs \ also shows a high mass-to-light ratio, which indicates that it is highly dark matter-dominated. The metal-poor conditions of DGSAT I may have enhanced the formation of massive stars, while at the same time, additional mechanisms are needed to prevent iron-rich yields from being recycled into stars. These results suggest that some ultra-diffuse galaxies could have experienced chemical enrichment episodes similar to the first building blocks of galaxies.

\end{abstract}



\section{Introduction}

The formation history of a galaxy leaves a permanent footprint on its chemical composition. In particular, the mass ratios of alpha-elements such as magnesium to iron ([Mg/Fe]) trace the time-scales for star formation, as these elements are produced by stars with different life-times \citep[e.g.][]{Worthey92,Thomas99,Thomas05,Pipino04}. Both giant elliptical galaxies \citep[e.g.][]{dlr11,Conroy14,LB14,McDermid15,MN18} and ultra-faint dwarfs \citep[e.g.][]{Norris10,Vargas13,Frebel14,Frebel16,Ishigaki14,Roederer16} are enriched in [Mg/Fe] by factors of up to 4 compared to the Milky Way. 

Recently, a population of galaxies with the sizes of giants but the luminosities of dwarfs has received considerable attention \citep[e.g.][]{Merritt14,Pieter15,koda15,Mihos15,Yagi16,MP18,Zaritsky18}. The nature and origins of these ultra-diffuse galaxies (UDGs) are under debate \citep{Amorisco16,dc17,Chan17,Rong17,Busola18,Bennet18,Carleton18,Ogiya18}, and it has been suggested that they may be analogs of the primordial stellar systems that built up the halos of more luminous galaxies \citep{Peng16}, with high levels of alpha elements.

There are indications that some UDGs are unusually metal-poor but detailed spectroscopic studies are still scarce due to the low-surface brightness of these objects \citep{Kadowaki17,Gu18,Tomas18,Anna18,Emsellem18,Fensch18}, which makes them observationally challenging even for 10-meter class telescopes. Based on photometric measurements, \citet{Viraj18} measured the stellar population properties of two UDGs. While one of these UDGs, VCC\,1287, turned out to be old and metal-poor as expected for its stellar mass, the other UDG in their sample,  \dgs, showed bluer optical colors \citep[see Figure 2 in][]{Viraj18}, reflecting the presence of young and/or metal-poor stellar populations. However, the lack of spectroscopic data prevented a more detailed stellar population analysis.

This relatively blue object DGSAT~I was first identified by \citet{MD16}, and it is particularly interesting because it apparently resides in a low density environment, at least compared to the densities where UDGs are typically found \citep[e.g.][]{vdB16,Roman17}. \dgs \ is a UDG likely associated with the outskirts of the  Pisces-Perseus supercluster  \citet{MD16}, with a central surface brightness of $\mu_V = 24.8$ mag arcsec$^{-2}$ and a total $V$-band luminosity of $L_V=2.7 \times 10^8$ $L_\odot$. It has an optical $(V-I)$ color of 1.0, and size of \re$=4.7$ kpc in the $I$-band at a distance of 78 Mpc \citep{MD16,Viraj18}. The estimated stellar mass of DGSAT~I is 4.8$\times10^8$ \msun, and it shows little sign of cold gas or current star formation \citep{Papastergis17}. 

The low-density environment where \dgs \ resides offers an opportunity to study the secular evolution of this type of object, and therefore, to understand different formation paths for UDGs without confounding environmental effects. In this paper, we make use of integral field spectroscopic data from the Keck Cosmic Web Imager (KCWI) to study in detail the stellar population properties of \dgs, as a representative example of a relatively isolated UDG. The layout is as follows: in Section~\ref{sec:data} we present the data. In Sections~\ref{sec:kine} and~\ref{sec:pops} we describe the kinematic and stellar populations analysis, respectively. The main results of this paper are discussed in Section~\ref{sec:discu}, and finally in Section~\ref{sec:sum} we summarize our main conclusions.

\section{Data}~\label{sec:data}
\begin{figure}
\begin{center}
\includegraphics[width=8.5cm]{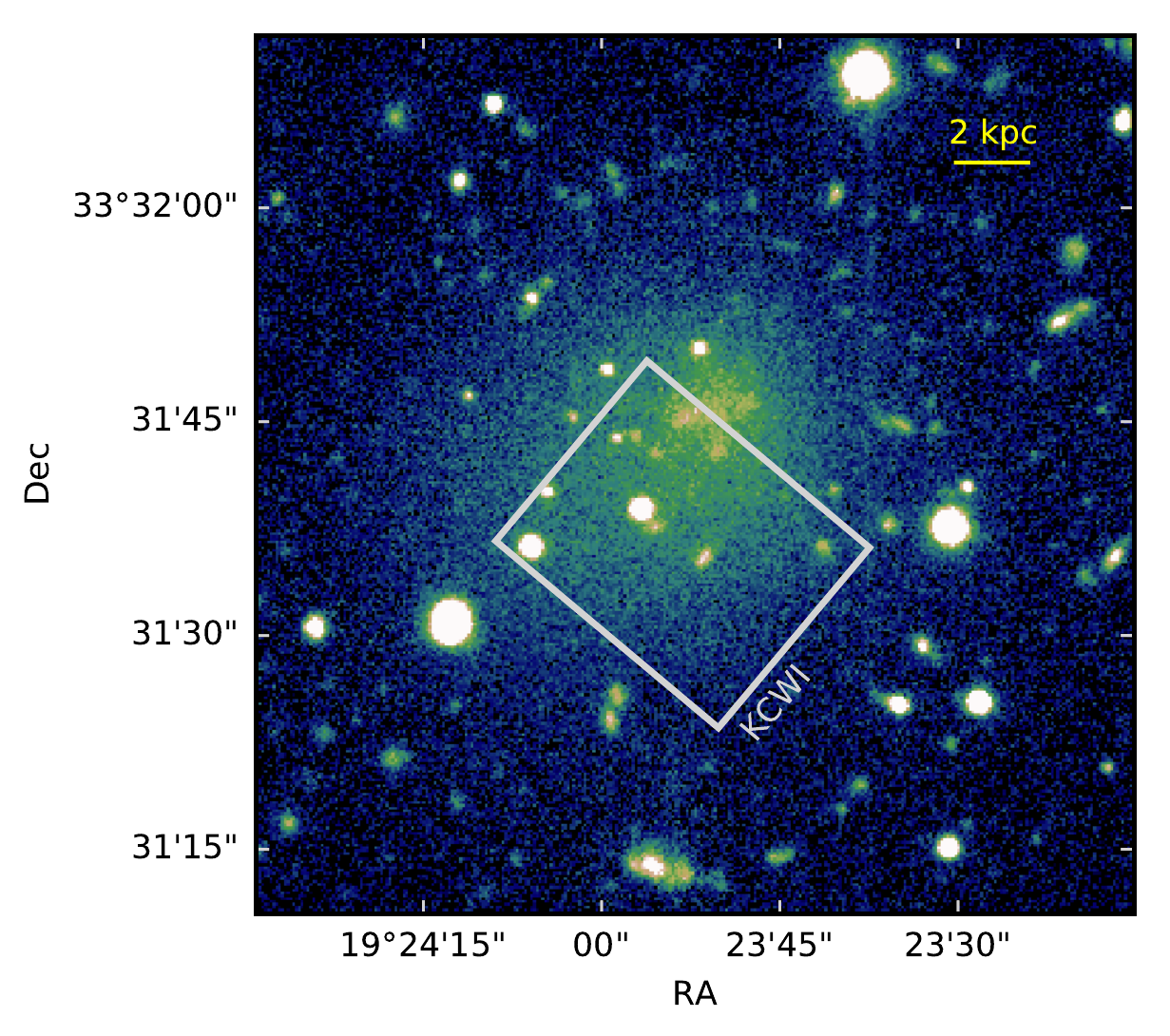}
\end{center}
\caption{Archival  Subaru/Suprime-Cam  $V$-band image of \dgs\ with the KCWI field-of-view marked in white. The galaxy photometric centre is marked with a cross. For reference, the yellow horizontal line has a length of 2 kpc.}
\label{fig:fov}
\end{figure}

Spectroscopic data were obtained through the Keck Cosmic Web Imager \citep{kcwi} integral field unit at Keck II Telescope, on 2017 Sept. 18,19 (PI Duncan Forbes) and Oct. 18,19 (PI Jean Brodie). We used the BM grating with the medium slicer, which gave us a spectral resolution of $R=4000$ over a field-of-view of 16.5" $\times$ 20.4" and a plate scale of $\sim$0.3 arcsec/spaxel. We covered a spectral range from $\lambda=4555$ to $\lambda=5370$ in the
rest frame. The total usable integration time was 7.25 hours (on source). We show in Figure~\ref{fig:fov} the $V$-band image of \dgs \ \citep{MD16}, with the KCWI field-of-view superimposed.

Data were reduced using the \textsc{kderp} KCWI pipeline\footnote{https://github.com/Keck-DataReductionPipelines/KcwiDRP}. It performs a standard data reduction process that covers a basic CCD reduction (bias and over-scan subtraction, gain-correction, trimming and cosmic ray removal), flat-fielding, geometric solution, profile correction and atmospheric correction. Additionally, it performs a (relative) flux-calibration using standard stars.

Background objects were masked before collapsing the KCWI field-of-view into a single spectrum: in particular, the two brightest stars visible in Figure~\ref{fig:fov}, along with a more distant galaxy only detected in emission (the extended object close to the center). In order to ensure a clean extraction of the \dgs \ spectrum, the masking process was done after collapsing all the KCWI exposures into a single image. In this way we maximized the signal-to-noise of the detection image, making sure we detected all possible objects contributing to the final spectrum. For safety, all masks extend well beyond the sources, where no more contaminated light is detected. This process was done manually in order to prevent any contamination in the spectrum of \dgs. Note also that the effective field of view of KCWI is slightly narrower than the nominal 16"x20" and thus, the brightest stars potentially affecting our results are barely included. No signs of additional objects were detected in the collapsed 2D cube. Due to the extremely low surface brightness of \dgs, we collapsed all the spaxels within the KCWI effective field of view (an aperture equivalent to $\sim 6\times7$~kpc assuming a 0.38 kpc/arcsec scale) into a single spectrum, reaching a signal-to-noise ratio of 30 \AA$^{-1}$.

\section{Kinematics}~\label{sec:kine}

Given the low surface brightness of \dgs, we carried out a more careful sky subtraction than that provided by the KCWI pipeline. We made use of pPXF \citep{ppxf,Cappellari17} to fit simultaneously both sky emission and galaxy stellar absorption. This technique, particularly well suited for sky-dominated data \citep[e.g.][]{Weijmans09}, allowed us to minimize the residuals in our final spectra. Since a proper sky subtraction is necessary for a reliable analysis, in Appendix~\ref{sec:radii} we tested the robustness of our results against changes in the sky subtraction process, finding no significant differences.

We fed pPXF with the PEGASE-HR stellar population synthesis models \citep{pegase,elodie}, which have a nominal resolution of R=10000, in order to measure the potentially low velocity dispersion of \dgs. We found a systemic velocity $V_\mathrm{sys} = 5439 \pm 8$ \kms and velocity dispersion $\sigma = 56 \pm 10$ \kms. The latter implies a mass of $M_{\rm dyn} = (1.3 \pm 0.5) \times 10^{10}$\msun\ within a deprojected half-light radius of 5.8~kpc \citep{Wolf10}, and an $I$-band mass-to-light ratio of $M/L_I = 71 \pm 25$. DGSAT~I is thus highly dark matter dominated in its central regions. The dynamical mass for \dgs was calculated at the half-light radius: a standard approach that is insensitive to orbital anisotropy. The specific formula used is $M_{\rm dyn}(< r_{1/2}) \simeq 9.3 \times 10^5 \sigma^2 R_{\rm e}$ \citep{Wolf10}, where $\sigma$ is in km~s$^{-1}$, $R_{\rm e}$ is circularized and in kpc, and $M_{\rm dyn}$ is in $M_\odot$.

In order to test the robustness of this result, we measured $V_\mathrm{sys}$ and $\sigma$ in three radial apertures (see Appendix~\ref{sec:radii}). We consistently recover a velocity dispersion value of 55 \kms, with a standard deviation of 7~\kms\ among these three radial apertures, in agreement with the integrated $\sigma$ measurement quoted in the main text. Since \dgs\ was not centered in the field-of-view of KCWI (see Figure~\ref{fig:fov}), the radial measurements of $V_\mathrm{sys}$ are sensitive to the rotation curve of this galaxy. We find slightly higher mean velocities in the outskirts ($V_\mathrm{sys}^{out} = 5445$ \kms) than in the center ($V_\mathrm{sys}^{cen} = 5434$ \kms). However, the large 1-$\sigma$ uncertainty, typically around 8 \kms, prevents us from any confident claim of rotation in \dgs. The signal-to-noise of the radial apertures was lower ($\sim20$) than that of the integrated spectrum. Although it is enough to confidently measure both $V_\mathrm{sys}$ and $\sigma$, it makes the radial analysis even more difficult. Its near-circular shape \citep{MD16} further complicates the identification of a clear rotation axis.

\section{Stellar population analysis}~\label{sec:pops}

\subsection{Star formation history}
We derived the star formation history (SFH) of \dgs\ by fitting our spectrum with STECKMAP \citep{Ocvirk06,Ocvirk06b} plus, as for the kinematics, the PEGASE-HR set of templates. In short, STECKMAP finds the best-fitting linear combination of single stellar populations models that reproduce a given spectrum. Recovering the SFH from an observed spectrum is an ill-defined problem, and therefore regularization is needed. The level of regularization imposed by STECKMAP is tuned via the dimensionless $\mu_x$ and $\mu_Z$ parameters, which effectively behave as Gaussian priors on the SFH and on the age--metallicity relation, respectively\citep{Ocvirk06b}. It is important to note here that age information is not only encoded in Balmer lines, but is also distributed along the whole optical wavelength range \citep{Ocvirk06}. We assumed $\mu_x=\mu_Z=10$, imposed by the signal-to-noise of the data \citep[see e.g.][]{Pat11} However, there is not a unique way of varying the amount of regularization and thus, the recovered SFH of \dgs\ may slightly depend on the adopted values for $\mu_x$ and $\mu_Z$. Appendix~\ref{sec:regu} demonstrates how our conclusions do not depend on the assumed  $\mu_x$ and $\mu_Z$ values.

\begin{figure}
\begin{center}
\includegraphics[width=8.5cm]{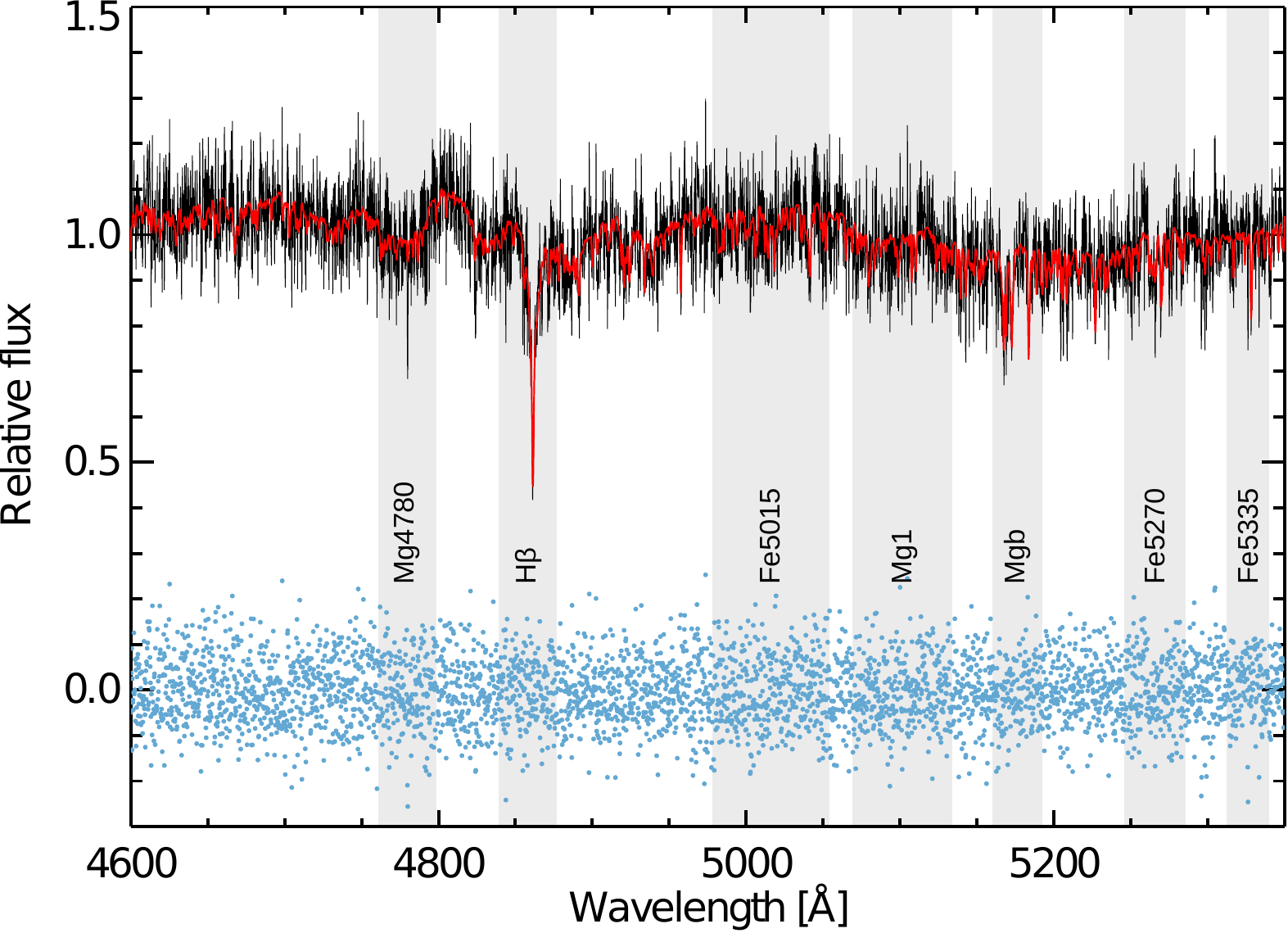}
\end{center}
\caption{The KCWI (rest-frame) spectrum of \dgs\ is shown in black, along with the best-fitting model derived with STECKMAP (red), and the residuals (blue). The spectrum was normalized so its median value was equal to one. The most prominent spectral features are clearly visible (vertical grey bands).}
\label{fig:spec}
\end{figure}

The best-fitting model and residuals are shown in Figure~\ref{fig:spec}, where the most prominent absorption features are clearly noticeable. \dgs\ shows an extended SFH, with a mass-weighted age of 8.1 $\pm$ 0.4 Gyr, and luminosity-weighted age of 2.8 $\pm$ 0.5 Gyr. It took $\sim 5$ Gyr for \dgs\ to form 50\% of its stellar mass. This result is in agreement with the photometric study of \citet{Viraj18}, where they estimated a luminosity-weighted age of $\sim$ 3 Gyr.

\begin{figure}
\begin{center}
\includegraphics[width=8.5cm]{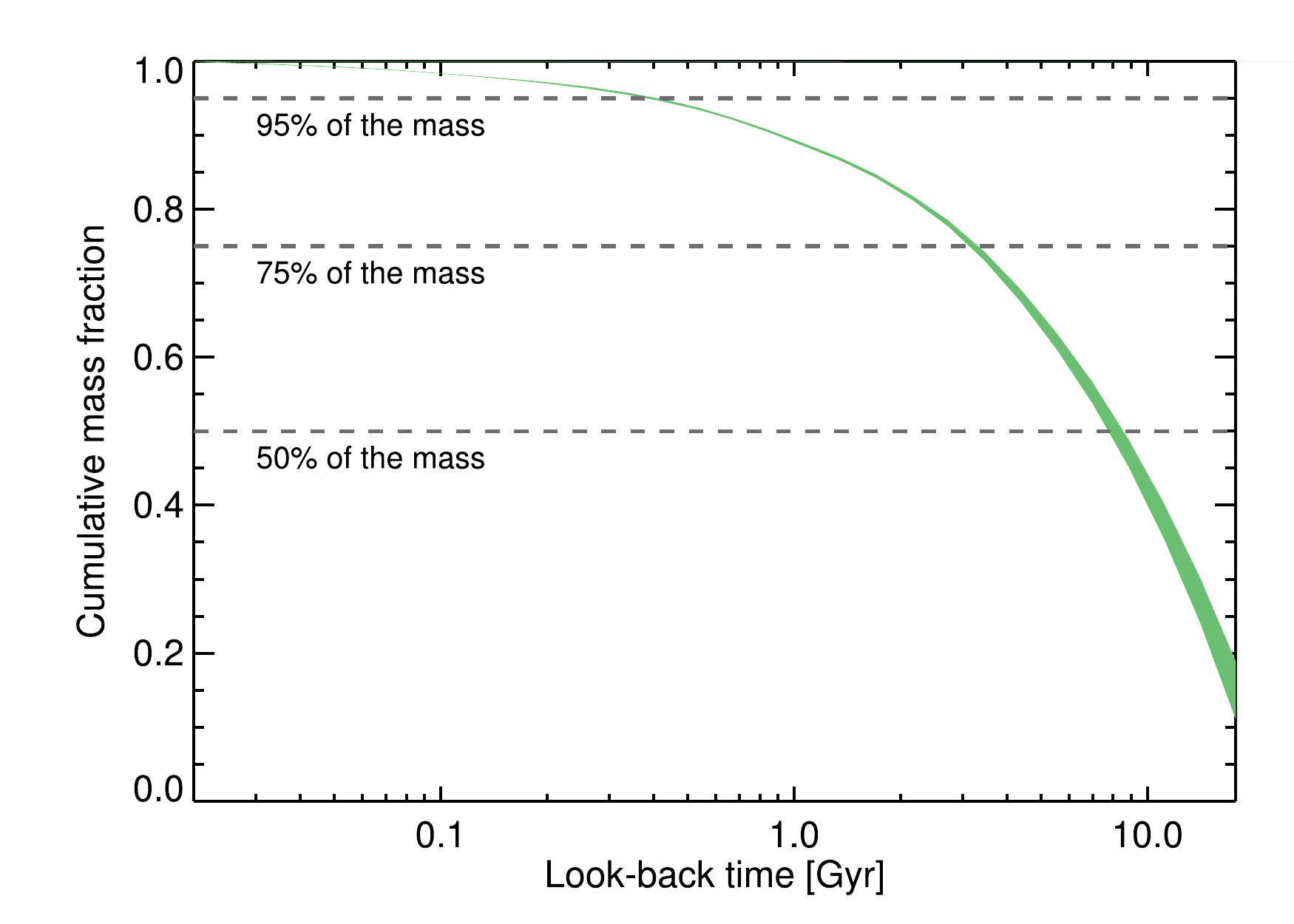}
\end{center}
\caption{Cumulative SFH of \dgs. Horizontal dashed lines indicate when this object had already formed 95\%, 75\%, and 50\% of its stellar mass, respectively. The width of the line corresponds to the estimated 1$\sigma$ uncertainty.}
\label{fig:sfh}
\end{figure}

In addition, we also measured the SFH of \dgs\ using a regularized pPXF solution \citep{Cappellari17}. This minimization scheme is different than the STECKMAP implementation, and it allows for multiple metallicities per age bin. Thus, a metallicity spread at a given age should be captured by pPXF. With this additional method, we recover a luminosity-weighted age of 3.5 Gyr, and a mass-weighted value of 7.9 Gyr, in excellent agreement with the STECKMAP analysis. Similar results were found using STARLIGHT\citep{starlight}. In order to quantify how much an extended SFH improves the fit over a single stellar population model, we compare the residuals to those expected from a Gaussian distribution with mean $\mu=0$ (i.e. age = 3.5 Gyr) and variance $\sigma=1$. A Kolmogorov--Smirnov test indicates that an extended star formation is rejected at a level of $0.37\sigma$, while the single-age population is strongly disfavored at a $3.9\sigma$ level. This test, along with the consistency among different algorithms, shows that \dgs\ experienced an extended SFH. The cumulative SFH of \dgs\ as measured from its optical spectrum is shown in Fig.~\ref{fig:sfh}.

Qualitatively, the young luminosity-weighted ages measured in \dgs \ using both STECKMAP, STARLIGHT, and pPXF suggest a long-lasting star formation activity in this object. Since even low-mass halos in the local Universe seem to host a fraction of very old stars \citep{Weisz11}, it is expected that the star formation in \dgs \ roughly lasted from the formation of these first stars in the early Universe until, at least, 3 Gyr ago.

\subsection{Chemical composition}

\begin{figure*}
\begin{center}
\includegraphics[width=15cm]{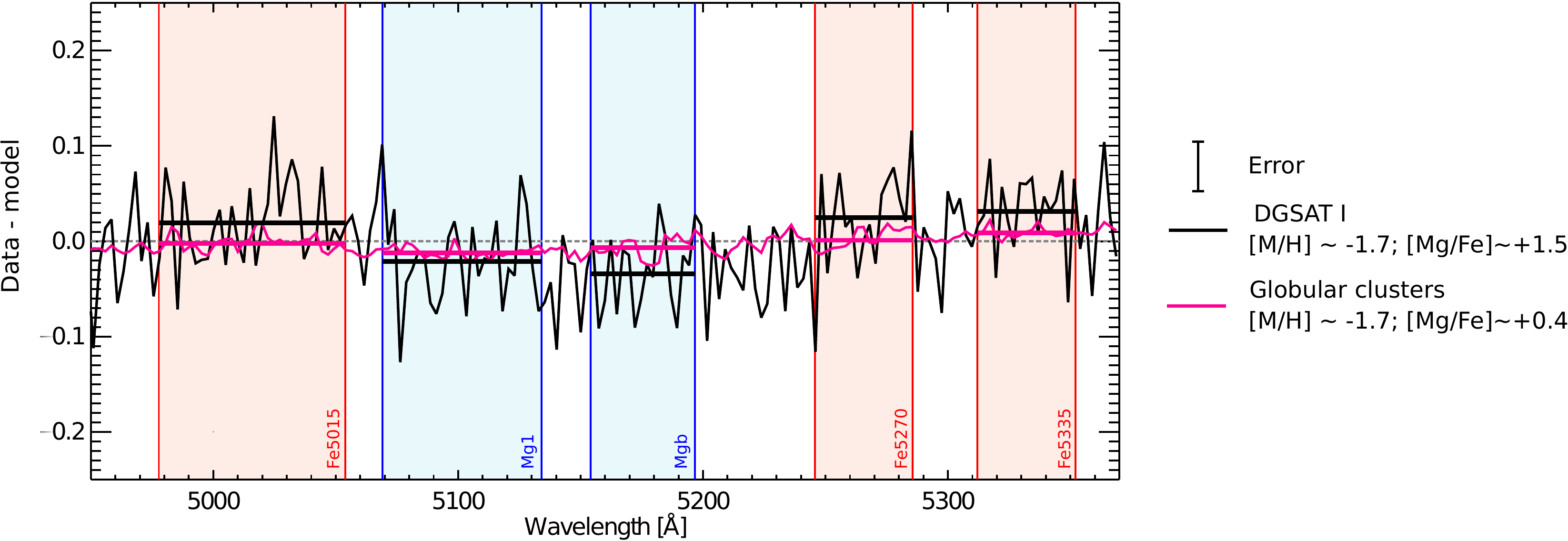} 
\end{center}
\caption{Empirical assessment of the abundance pattern of \dgs. In black we show the residuals between the observed spectrum of \dgs\ and a metal-poor ([M/H] = $-1.7$), $\alpha$-enhanced ([Mg/Fe] = $+0.4$) stellar population model at rest-frame. Vertical colored regions indicate the index band-pass definition. While magnesium lines (in blue) are under-predicted by the model, iron lines (in red) behave in the opposite way, showing that the stellar populations of \dgs \ are more $\alpha$-enhanced than the model itself. For comparison, we present the same ratio but for metal-poor globular clusters in the Milky Way (magenta) \citep{Usher17}. In this case, the observed spectrum is well represented by the stellar population model, without any additional Mg-enhancement. The typical 1-$\sigma$ uncertainty for  \dgs \ spectrally binned data is shown as an error bar. Data and models were convolved with a 2.5 \AA-width boxcar kernel.}
\label{fig:empi}
\end{figure*}

\dgs's optical spectrum (Figure~\ref{fig:empi}) reveals the peculiar chemical composition of this object. When an $\alpha$-enhanced ([Mg/Fe] = +0.4), metal-poor ([M/H] = $-1.7$) stellar population model (red line in Fig.~\ref{fig:spec}) is subtracted from the observed spectrum, the residuals associated with iron-sensitive and magnesium-sensitive features behave in the opposite manner. Magnesium lines are under-predicted by the model (negative residuals), while iron lines are over-predicted (positive residuals). For comparison, Figure~\ref{fig:empi} also shows the residuals of metal-poor, $\alpha$-enhanced globular clusters \citep{Usher17}. The observed spectrum of \dgs \ and that of the globular clusters were divided by a constant factor so their median is equal to one. Note that best-fitting models account for possible systematics in the (relative) flux calibration of the data (see Appendix~\ref{sec:fluxcal}).

\begin{figure}
\begin{center}
\includegraphics[height=5.5cm]{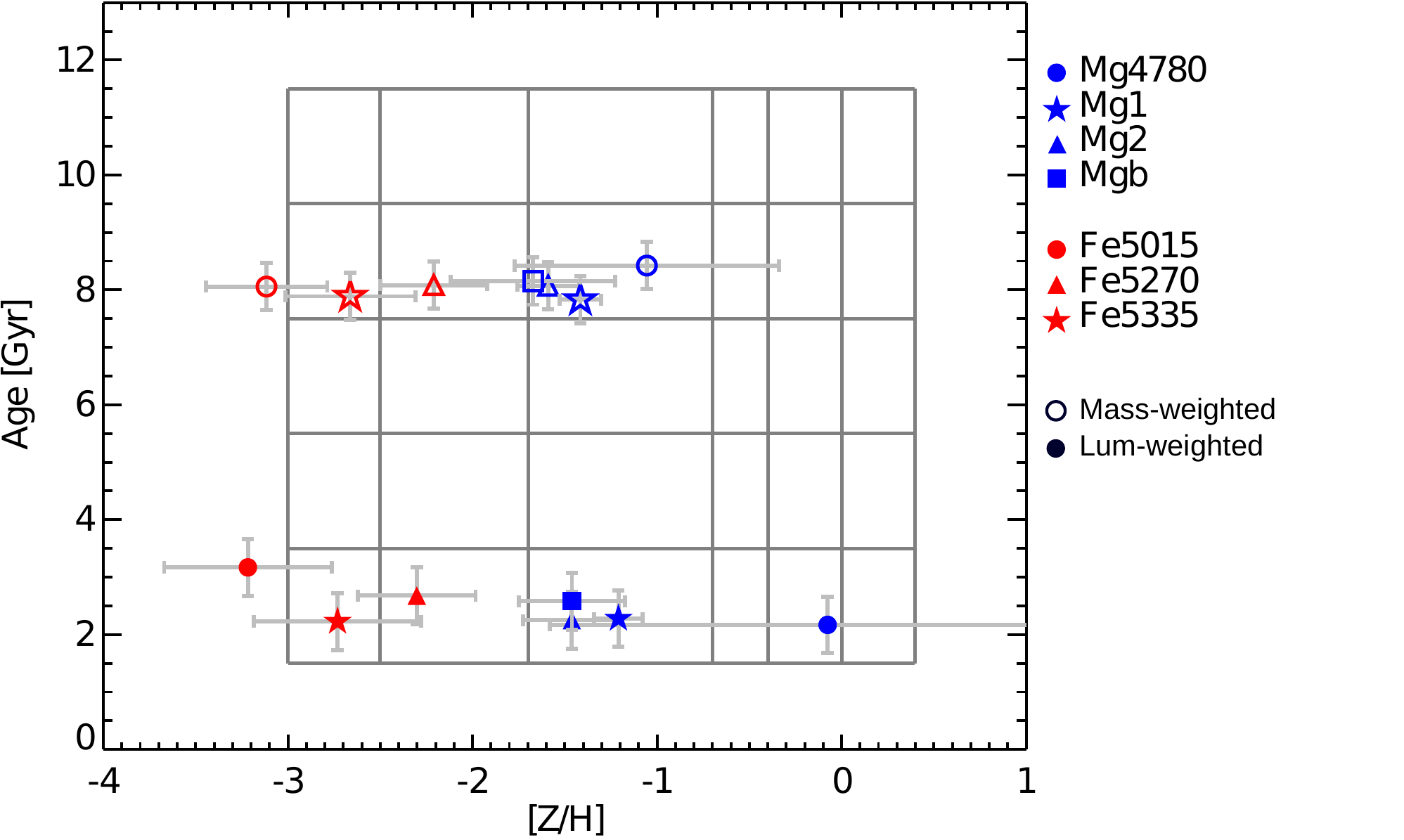} 
\end{center}
\caption{Line-strength analysis. The best-fitting metallicities based on magnesium lines are shown with blue symbols, while metallicities measured from iron lines are shown in red. Open and filled symbols result from assuming the mass-weighted (8.1 Gyr) and the luminosity-weighted (2.8 Gyr) ages measured with STECKMAP, respectively. For clarity, ages are vertically shifted according to the 1-$\sigma$ uncertainties (vertical error bars). Independently of the assumed age, magnesium-sensitive lines indicate a much higher metallicity than that derived from iron-sensitive features, clearly revealing the abnormal [Mg/Fe] enhancement of \dgs. Note that the stellar population model at these metallicities is already $\alpha$-enhanced.}
\label{fig:indices}
\end{figure}

In order to quantify the chemical properties of \dgs\ we followed the standard line-strength analysis. We partially broke the age--metallicity degeneracy by fixing the age to those (luminosity- and mass-weighted) obtained with STECKMAP. The best-fitting metallicity was then derived by comparing the equivalent width of metallicity-sensitive lines in our spectrum to the PEGASE-HR model predictions (at fixed age), as show in Fig.~\ref{fig:indices}. The typical error in the line-strength measurements was $\sim0.2$\AA, which leads to the uncertainties shown in Fig.~\ref{fig:indices}.

We found a metallicity value of [M/H] = $-1.8 \pm 0.4$ dex assuming a mass-weighted age of 8.1 Gyr, as expected for a low-$\sigma$ galaxy. However, metallicity measurements based on magnesium spectral features ([M/H]$_\mathrm{Mg} = -1.0 \pm 0.5$) are higher than the metallicities estimated from iron lines ([M/H]$_\mathrm{Fe} = -2.7 \pm 0.3$), as shown in Figure~\ref{fig:indices}. Note that PEGASE-HR stellar population models are based on solar neighborhood stars, which are magnesium-rich in the low-metallicity regime. Hence, given the low-metallicity of \dgs, iron- and magnesium-based metallicities were in practice measured with respect to stellar population models with [Mg/Fe] $\sim+0.4$. This implies that the stellar populations of \dgs\ have an [Mg/Fe] enhancement larger than the metal-poor stars in the solar neighborhood, i.e., larger than +0.4 dex. In the Appendix (Fig.~\ref{fig:flux_index}) we test the robustness of the measured [Mg/Fe] against flux calibration issues.

Both the empirical analysis of the \dgs\ spectrum and the difference between magnesium- and iron-based metallicities highlight the abnormal chemical enrichment of this UDG. A zero-order estimate of its [Mg/Fe] ratio can be done using the proxy $Z_\mathrm{Mg} -$ $Z_\mathrm{Fe}$ \citep{Vazdekis15}, and leads to a value of [Mg/Fe] $\backsimeq + 1.0$ dex. This estimate should be interpreted as a lower limit, since it does not take into account the decreasing sensitivity to [Mg/Fe] with decreasing metallicity \citep{Vazdekis15}.  To fully quantify the [Mg/Fe] ratio of \dgs, we made use of the new set of MILES stellar population models \citep{miles,Vazdekis15}. This set of models consistently treat the effect of the [$\alpha$/Fe] ratio across a wide range of metallicities, from [M/H] = $-2.27$ to [M/H] = +0.26. The MILES resolution is, however, lower than our data. We therefore extrapolated the line-strength MILES predictions to match the resolution of the KCWI spectrograph. This correction, typically $\sim100$ times smaller than the error in the equivalent widths, is negligible compared to the intrinsic uncertainty of the data. 

\begin{figure}
\begin{center}
\includegraphics[height=5.3cm]{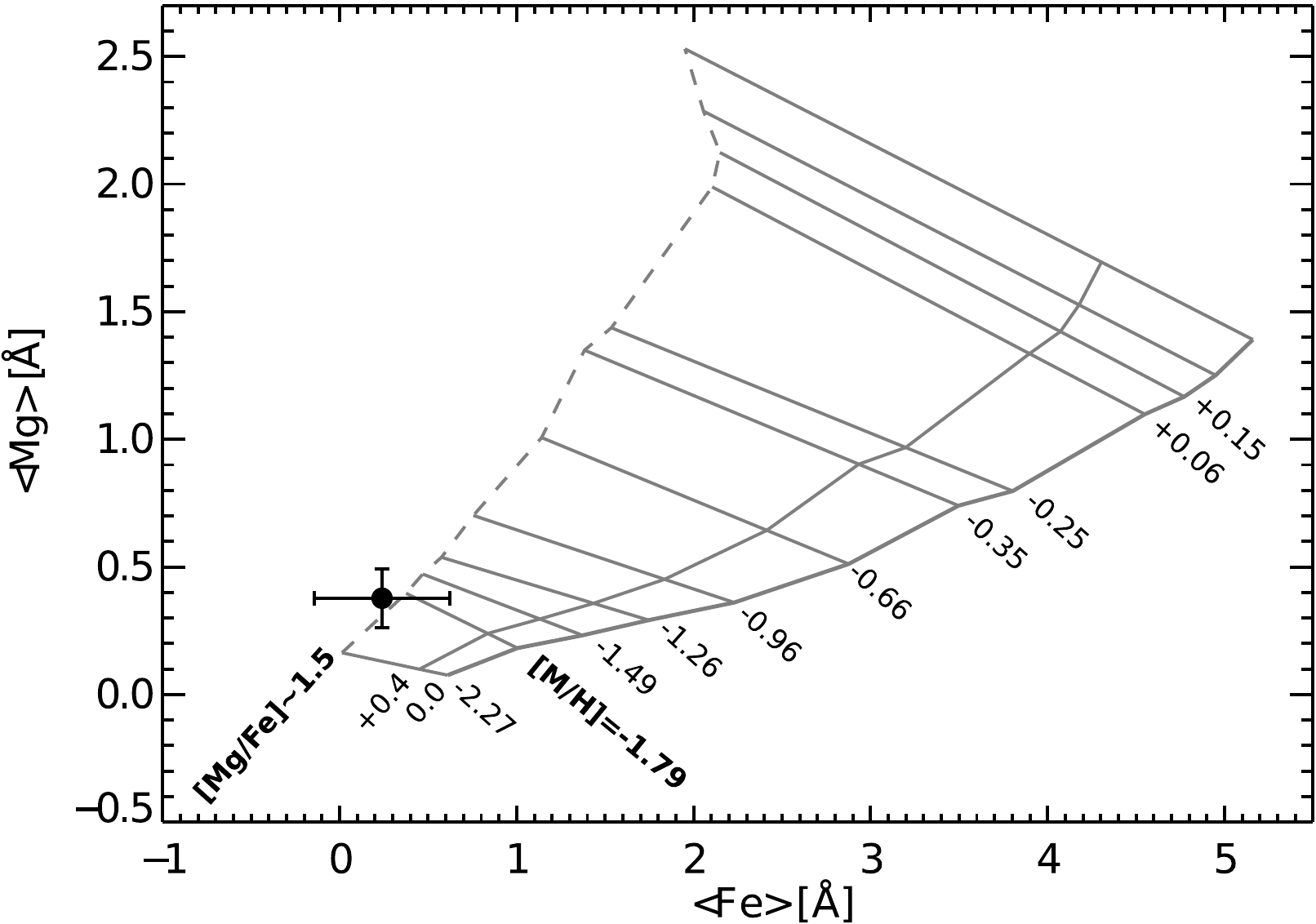} 
\end{center}
\caption{The index--index grid based on the MILES models shows that the stellar populations of \dgs\ are metal-poor ([M/H]$\sim -1.8$) and extremely [Mg/Fe] enhanced ([Mg/Fe] $\sim+1.5$). The lower sensitivity of the [Mg/Fe] at low metallicities is also clear, as the same difference in [Mg/Fe] implies a weaker variation in the line-strength indices than in the metal-rich regime.}
\label{fig:grid}
\end{figure}

The [Mg/Fe] measurement was done by fitting the average of magnesium and iron lines equivalent widths, as shown in Fig.~\ref{fig:grid}, assuming a luminosity weighted age of 3.5 Gyr for the MILES $\alpha$-enhanced models. We fixed the slope of the stellar initial mass function to that of the Milky Way \citep{Kroupa}. The measured value for \dgs \ corresponds to an iron metallicity of [Fe/H] $\backsimeq -2.8 \pm 0.3$, and an extreme magnesium enhancement of [Mg/Fe] $\backsimeq+1.5\pm0.5$. These values should however be interpreted with caution given the extreme extrapolations of the models predictions. The relative comparison shown in Fig.~\ref{fig:indices} is a more robust evidence of the peculiar abundance pattern of \dgs. The agreement between the results obtained using PEGASE-HR and MILES models indicates that the extreme [Mg/Fe] ratio of \dgs\ is not related to systematics in the stellar population modeling.

\section{Discussion}~\label{sec:discu}

The stellar population properties of \dgs \ appear to contradict one another. Under a canonical interpretation, a high [Mg/Fe] ratio is produced in short-lasting star formation events \citep{Thomas05}, which cease before Type Ia supernovae (SN~Ia) explode and pollute the inter-stellar medium (ISM) with iron-rich ejecta. However, the SFH of \dgs\ is very extended, lasting for more than 10 Gyr, which leaves plenty of time for SN~Ia to enrich the ISM with iron. 

Reconciling a long-lasting SFH with such a high [Mg/Fe] ratio suggests exotic processes regulating the chemical evolution of \dgs. Since magnesium is mainly produced in massive stars, it is natural to relate [Mg/Fe] enhancement to a stellar initial mass function (IMF) dominated by massive stars, i.e. a top-heavy IMF. Observationally, there is evidence suggesting that metal-poor galactic systems, as measured in this UDG, tend to favor the formation of massive stars \citep{Worthey92,Geha13,MN15}. Hence, a relatively flat IMF would naturally alleviate the tension between the formation time-scale and the chemical composition of \dgs, although it would also speed up its chemical enrichment. The metal-poor environment within \dgs\ may have also favored the formation of peculiar SNe whose ejecta are overly magnesium-enhanced \citep{Aoki07}. 

The lack of iron, however, is harder to reconcile with an extended SFH. It appears that \dgs\ efficiently retained the magnesium-rich yields of SN II, but it was unable to recycle iron-rich SN~Ia stellar ejecta. Such a selective mass loss could be consistent with a scenario where the large sizes of UDGs are due to supernovae-driven gas outflows, expanding both stellar and dark matter components \citep{dc17,Chan17}. Under this scenario, it would be possible that magnesium-rich yields were produced in a deeper potential well, before the galaxy expansion. In contrast, SN~Ia would have exploded in a much shallower halo, facilitating the escape of their iron-rich ejecta. Moreover, the cooling time of stellar ejecta increases with decreasing density and metallicity \citep{Martizzi15}, further delaying the recycling of iron-rich ejecta after the initial galaxy expansion.  

Alternatively, it could be argued that the low stellar density of \dgs\ may suppress the formation of binary stars and therefore, of SNe Ia. However, at fixed metallicity, the [Mg/Fe] ratio does not seem to increase with decreasing stellar density \citep{Shetrone01,Pritzl05}, suggesting that iron-rich stellar ejecta are recycled even in very low-mass galaxies \citep{Vargas13}. More extreme processes to explain the lack of iron in \dgs\ such as filamentary (Mg-rich) gas accretion from the cosmic web could also be possible \citep{Jorge17}, although it would require a constant cold accretion over a period of $\sim 10$ Gyr. Note also that this process is not unique to UDGs and would also apply to other (dwarf) galaxies where the chemical composition is known to be less extreme.

\begin{figure}
\begin{center}
\includegraphics[height=7.cm]{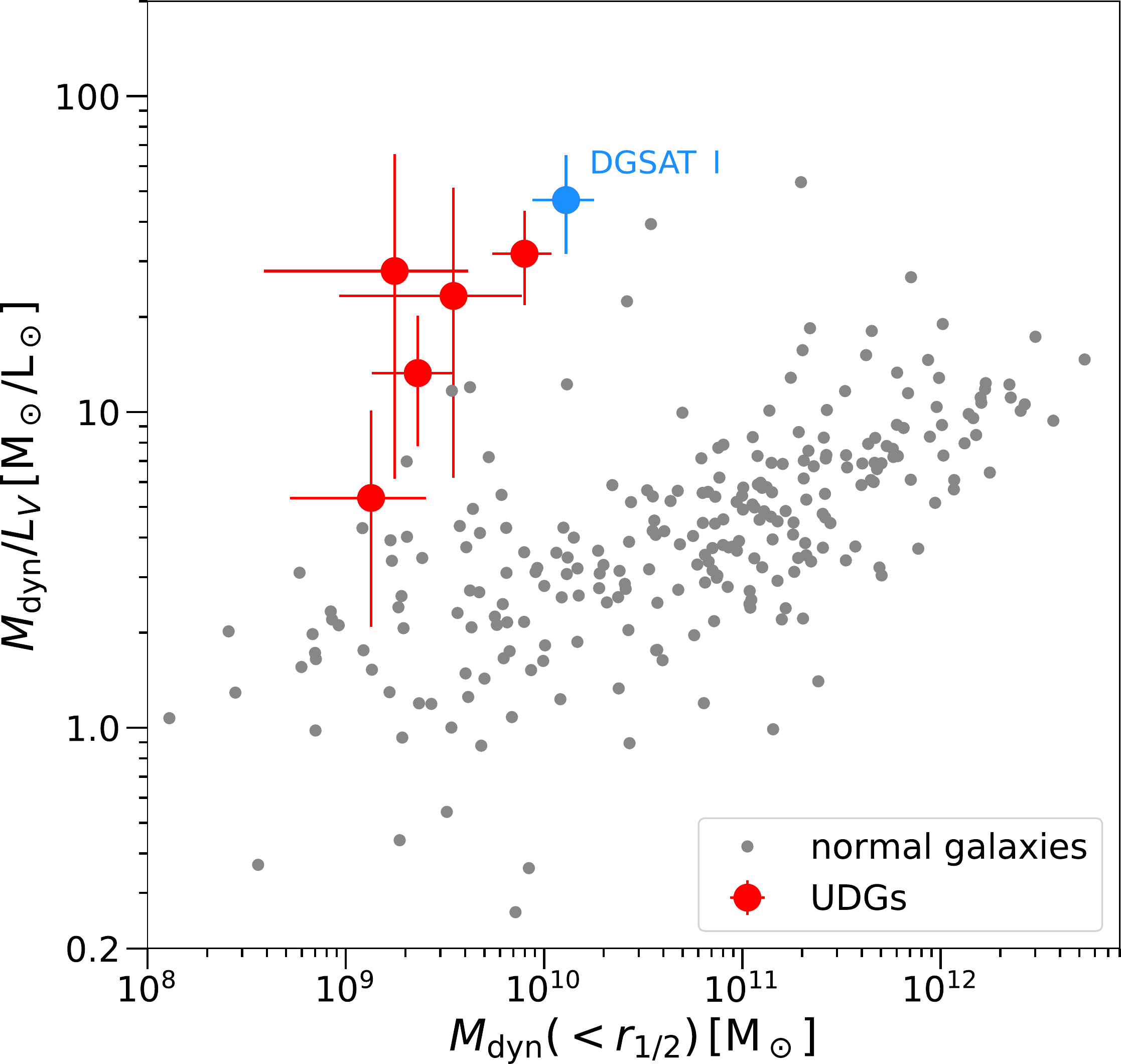} 
\includegraphics[height=6cm]{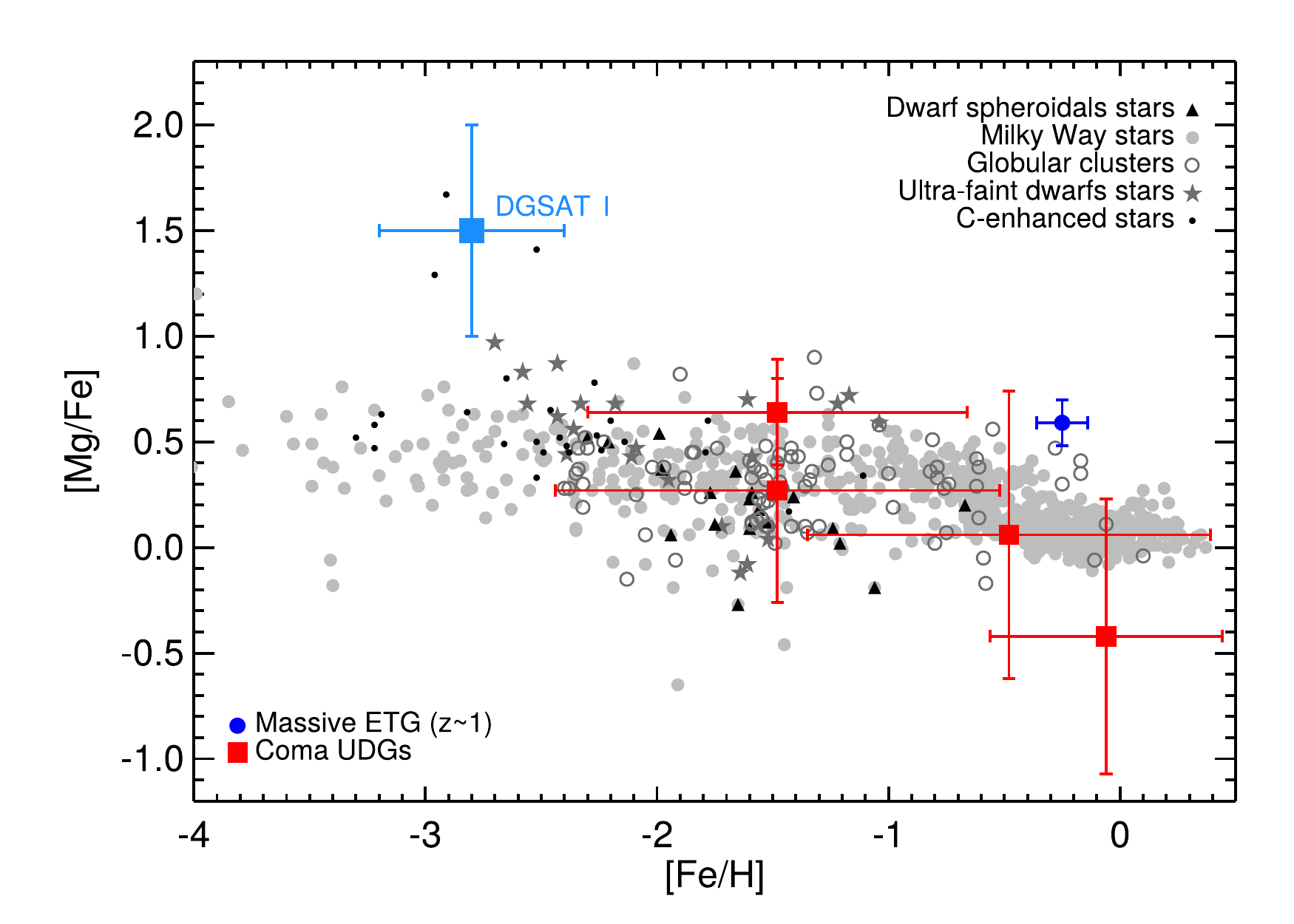} 
\end{center}
\caption{Scaling relations. In the upper panel we show the dynamical mass-to-light ratio vs.\ the dynamical mass, both within a half-light radius. \dgs\ and other UDGs (red points) are outliers relative to other objects (grey), with an excess of dark matter. In the bottom panel we show the [Mg/Fe] ratio vs.\ iron metallicity, for field stars and globular clusters \citep{Pritzl05} in the Milky Way and in satellite galaxies \citep{Shetrone01,Vargas13,Aoki07}. A massive early-type galaxy at high redshift is also shown \citep{Kriek16}, along with Coma UDGs \citep{Anna18}. They follow the Milky Way relation of increasing [Mg/Fe] at lower metallicities, with two UDGs resembling average halo stars. \dgs\ is more extreme, with lower metallicity and higher [Mg/Fe].}
\label{fig:scaling}
\end{figure}

Fig.~\ref{fig:scaling} shows how \dgs\ compares to the general population of galaxies. Similarly to some other UDGs in denser environments \citep{Mike16,Pieter16,Toloba18}, \dgs\ is heavily dark-matter dominated, deviating strongly from the scaling relation of normal galaxies. Its abnormal chemical composition is also clear from Fig.~\ref{fig:scaling}, where \dgs\ exhibits a [Mg/Fe] $\sim$ 1.0 dex higher than any other galaxy known so far. Interestingly, there is tentative evidence of a trend within the UDG population, with more metal-poor objects being more [Mg/Fe] enhanced, similarly to other low-mass galaxies \citep{Conselice03,Penny08,Anna18} and star in the Milky Way. Only a handful of resolved Milky Way stars \citep{Aoki07} present [Mg/Fe] values similar to that measured in \dgs. The similarity with these stars, likely descendants of the first core-collapse SNe \citep{Audouze95}, further supports the pristine character of \dgs. However, it is not yet understood how its stellar populations have evolved unaltered after 10 Gyr of star formation. Neither the metallicity of \dgs, nor its mass, nor its size alone would be enough to explain how this galaxy had an extended SFH while preserving a high [Mg/Fe] ratio. Therefore, the evolutionary history of \dgs\ is likely to be the differentiating factor. If so, it would put strong constraints on the formation of low-density cores within low-mass galaxies via supernovae feedback \citep[e.g.][]{Governato10,Pontzen12}, as this process should be then reflected in their chemical abundance patterns. 

\section{Summary and conclusions}~\label{sec:sum}

\dgs \ is a relatively isolated UDG with a measured velocity dispersion of $\sigma \sim56$ \kms, indicating a high dark matter-to-stellar mass ratio. Moreover, \dgs \ exhibits an extended SFH, in agreement with its blue colors. However, this long-lasting star formation time-scale is in apparent contradiction with its chemical properties, as \dgs \ hosts extremely magnesium-enhanced stellar populations. Reconciling such a high [Mg/Fe] abundance ratio, which seems to be a common feature of at least some UDGs in denser environments \citep{Tomas18,Anna18}, with an extended SFH requires an inefficient recycling process of Fe-rich SN\,Ia stellar ejecta, either due to the low density of \dgs, or as a consequence of the particular formation path of this object.

Regardless of the exact mechanism(s) regulating the chemical evolution of \dgs, its stellar populations may provide a window into the early Universe. The lack of iron is an indication of undeveloped chemical evolution, suggesting that the star formation conditions were similar to those of primeval galaxies. Spectroscopic observations of larger samples of UDGs \citep{Tomas18,Anna18} are needed in order to further investigate the connections between star formation, abundance patterns and formation pathways for this class of exotic objects \citep{Amorisco16,dc17,Chan17}, in particular for isolated UDGs.

\section*{Acknowledgments}

We acknowledge support from the National Science Foundation grants AST-1616598, AST-1616710, Marie Curie Global Fellowship, and from grants AYA2016-77237-C3-1-P and AYA2014-56795-P from the Spanish Ministry of Economy and Competitiveness (MINECO). AJR was supported as a Research Corporation for Science Advancement Cottrell Scholar. DAF and AFM thank the ARC for financial support via DP160101608.  also acknowledges financial support through the Postdoctoral Junior Leader Fellowship Programme from ”la Caixa” Banking Foundation (LCF/BQ/LI18/11630007). MES is
supported by the RFBR grant no. 18-02-00167 A. The data presented herein were obtained at the W. M. Keck Observatory, which is operated as a scientific partnership among the California Institute of Technology, the University of California and the National Aeronautics and Space Administration. The Observatory was made possible by the generous financial support of the W. M. Keck Foundation.  DMD acknowledges support by Sonderforschungsbereich (SFB) 881 ``The Milky Way System'' of the German Research Foundation (DFG) (sub-project A2). We would like to thank P.\ van Dokkum, R.\ Leaman, G.\ van de Ven, A.\ L.\ Frebel and S. Penny for their useful comments. The authors wish to recognize and acknowledge the very significant cultural role and reverence that the summit of Maunakea has always had within the indigenous Hawaiian community. We are most fortunate to have the opportunity to conduct observations from this mountain.




\bibliographystyle{mnras}
\bibliography{DGSAT} 


\appendix

\section{Robustness of the stellar population analysis} \label{sec:robust}

\subsection{Star formation histories} \label{sec:regu}
\begin{figure}
\begin{center}
\includegraphics[width=8cm]{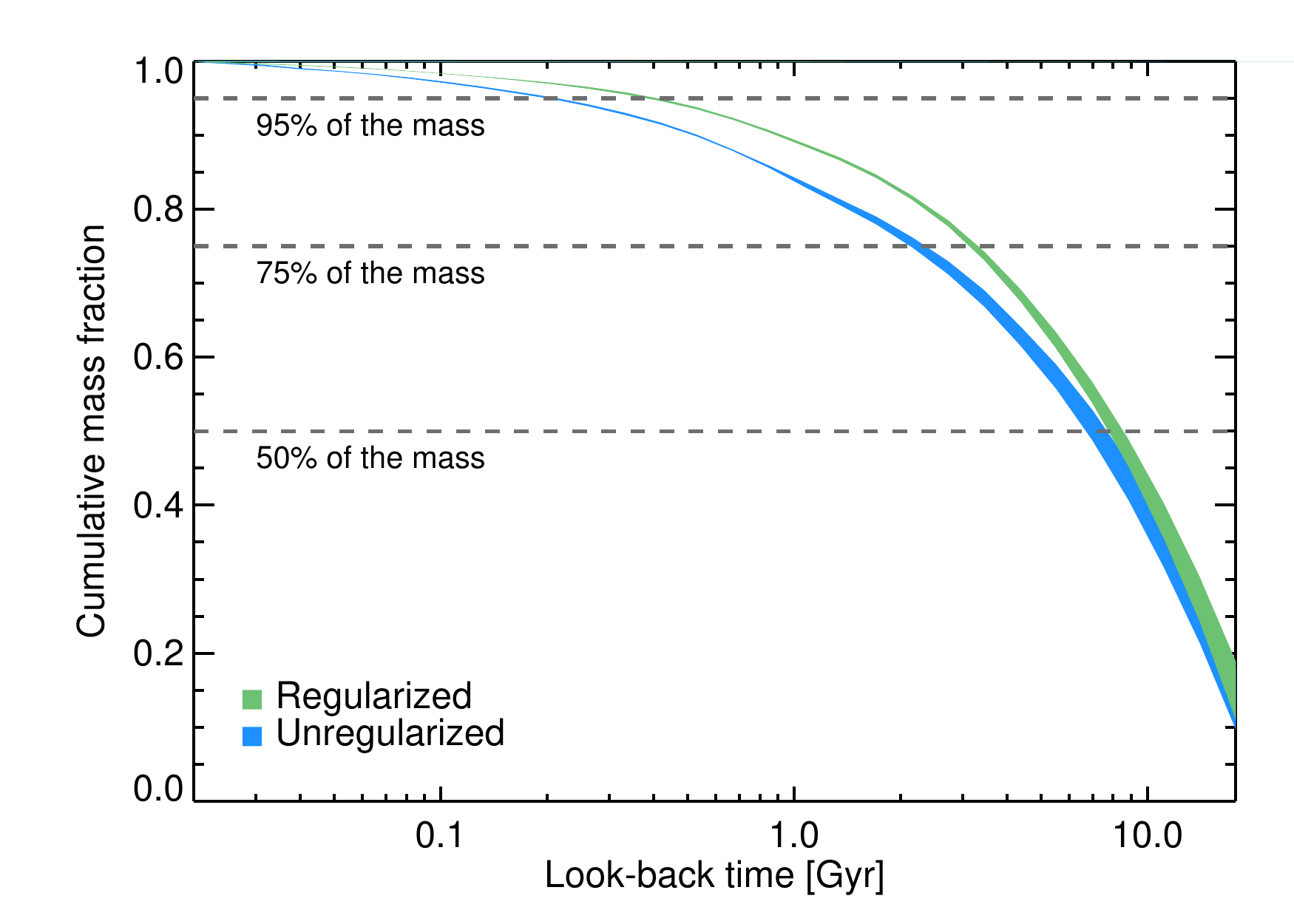}
\end{center}
\caption{The extended SFH of \dgs. In green we show the fiducial cumulative mass function for \dgs, assuming $\mu_x=\mu_Z=10$. Horizontal dashed lines mark when 50\%, 75\% and 95\% of the stellar mass was formed. In blue we show the cumulative mass function but leaving the STECKMAP solution almost un-regularized ($\mu_x=\mu_Z=10^{-4}$), which tends to favor even more extended star formation histories.Our choice of the regularization parameters is therefore not spuriously delaying the mass assembly of this object.}
\label{fig:regu}
\end{figure}

A key point in our analysis is the fact that \dgs\ exhibits an extended SFH. To investigate to what extent our choice of the regularization parameters affects the measured SFH, we repeated the STECKMAP analysis but assuming  $\mu_x=\mu_Z=10^{-4}$. In Fig.~\ref{fig:regu} we show the cumulative mass distribution of \dgs\ with ($\mu=10$) and without ($\mu=10^{-4}$) regularization. In both cases, the star formation lasts for more than 10 Gyr, appearing even more extended in the case without regularization. Hence, regularization is not responsible for the long-lasting SFH of \dgs. As an additional test, we also measured the SFH from previously published optical spectra \citep{MD16}. Although in this case the spectral coverage was not enough to measure the abundance pattern of \dgs, we obtained ages and metallicities in agreement with the KCWI data.

Finally, it is worth noting that a significant population of blue horizontal branch (BHB) stars in metal-poor objects could artificially bias measured ages towards younger values \citep{Cenarro08,Conroy18}, which may affect the recovered SFH of \dgs. This bias is introduced because BHB stars do not follow canonical isochrones, being hotter than expected for their luminosity and thus, they are not properly modelled by standard stellar population synthesis models. Effectively, this means that they can be misinterpreted as a young (hot) component. However, it has been shown that, using STECKMAP, the net effect of BHB stars is to resemble a burst of recent (age $\sim$ 0.1 Gyr) star formation \citep{Ocvirk10}. Similar results (i.e. a burst of young stars rather than a more extended star formation) have been found for similar algorithms \citep{Coelho09}, which, in practice, could be used to identify BHB stars in unresolved stellar populations \citep{Koleva08}. Hence, our measurement of an extended SFH in \dgs\ is not likely due to an inaccurate modeling of the horizontal branch morphology.  

\subsection{Line-strength analysis} \label{sec:radii}

Because of the unusually high [Mg/Fe] measurement, we have thoroughly looked for systematic effects on the data. With a relatively low signal-to-noise of 30 per \AA, metallicity and abundance pattern measurements based on individual spectral features are unreliable. However, as shown in Fig.~\ref{fig:indices}, all magnesium and iron line residuals have opposite behaviors. This is a crucial point, since different line-strengths have different band-passes and pseudo-continuum definitions. The fact that magnesium-based metallicities are $\sim 1$ dex higher than iron-based ones, independently of the index choice, shows that our [Mg/Fe] estimation is not driven by data systematics such as flux calibration issues.

\begin{figure}
\begin{center}
\includegraphics[width=8.5cm]{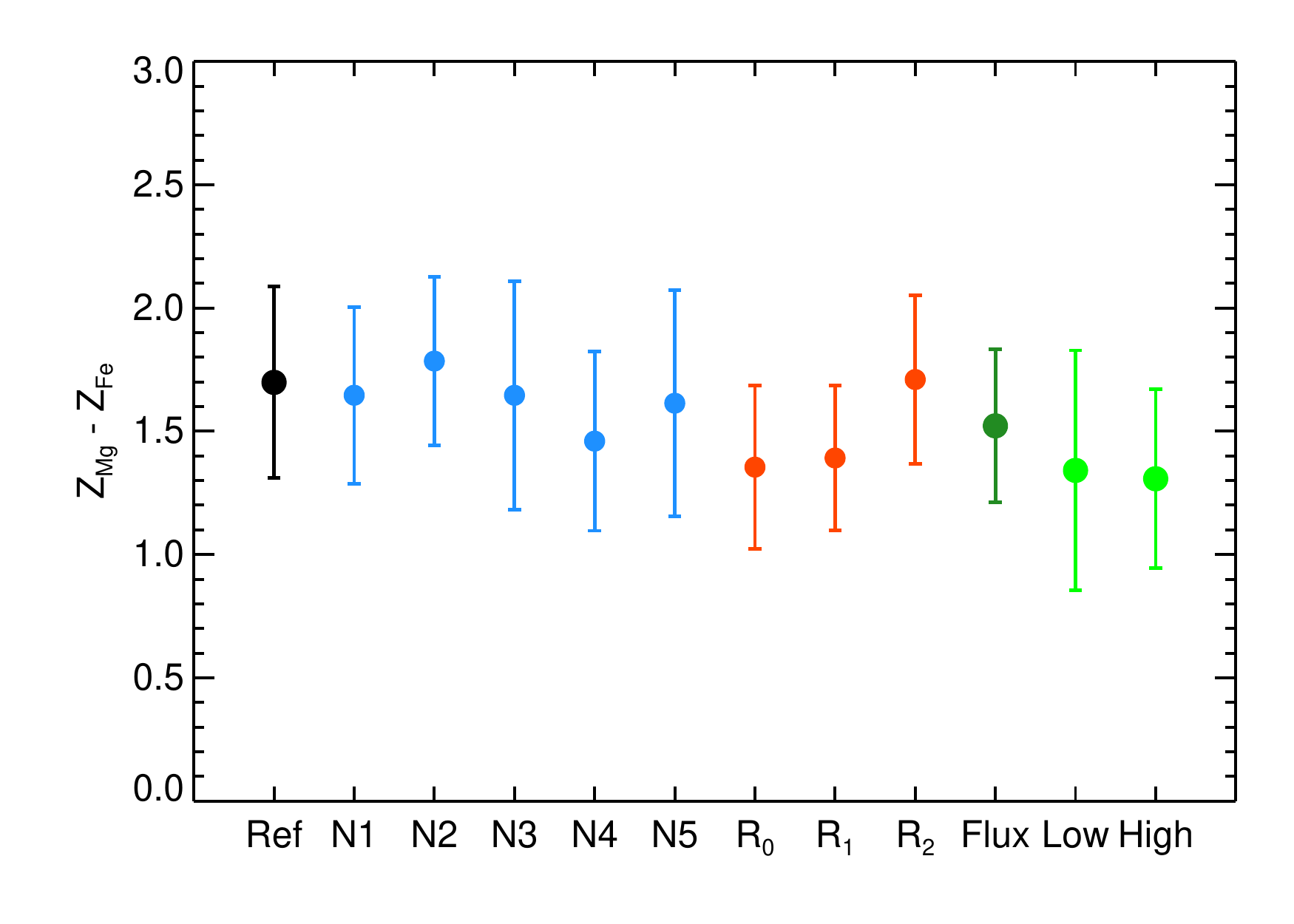}
\end{center}
\caption{Robustness of the [Mg/Fe] measurement. Differences between magnesium-based and iron-based metallicities (Z$_\mathrm{Mg}$ and Z$_\mathrm{Fe}$, respectively) for the tests described above. In black we show our reference value. In blue we show the five bootstrapping  realizations, where the measurement was done over 20 randomly selected frames out of the 36 individual exposures. In red, we show the radial profile, from the centre (R$_0$), to the outskirts (R$_2 \sim 2$ kpc). Dark green symbol shows the measurement after removing the index continua (see details below), while the two light green points are the result after varying the second-order polynomial correction to the continuum. {\it Low} corresponds to a low-order polynomial, while {\it high} indicates a higher order correction. The agreement between all these measurements indicates that systematics in our data are not likely driving the observed abundance pattern. Error bars correspond to 1-$\sigma$ uncertainties.}
\label{fig:tests}
\end{figure}

We made two extra tests to further assess the robustness of our [Mg/Fe] measurement. First, we randomly selected 20 of the 36 individual exposures of \dgs\ and then measured the stellar population properties only over this subset of the data. We repeated this process five times, making sure that our results are not driven by anomalous frames. However, we find that, irrespectively of the sub-sample of frames, magnesium-based and iron-based metallicities systematically depart from each other, as shown in Fig.~\ref{fig:tests}. We also repeated this test but stacking consecutive exposures, and found the same result. 

As a final test, we  attempted a radial analysis of our IFU data, defining 3 circular apertures (annuli) at $R_0=0$ kpc, $R_1\backsimeq0.8$ kpc, and $R_2\backsimeq2$ kpc from the center of the galaxy. If the [Mg/Fe] over-enhancement of \dgs\ were to due to small-scale variations in the flux calibration, [Mg/Fe] measurements should abruptly change across the detector. However, we find a rather constant magnesium-enhancement over the whole KCWI field-of-view. There is only marginal evidence of a slightly higher [Mg/Fe] ratio in the outskirts of \dgs, which would be in agreement with a standard outside-in formation scenario \citep{Pipino04}. Also, note that the metallicity does not significantly change  across the galaxy, although the large uncertainties in the metallicity measurement prevent us from obtaining a reliable radial gradient. That indicates that our measurements are not driven by the  off-center over-density in the North-West of DGSAT~I \citep{MD16}.

\begin{figure}
\begin{center}
\includegraphics[width=8.5cm]{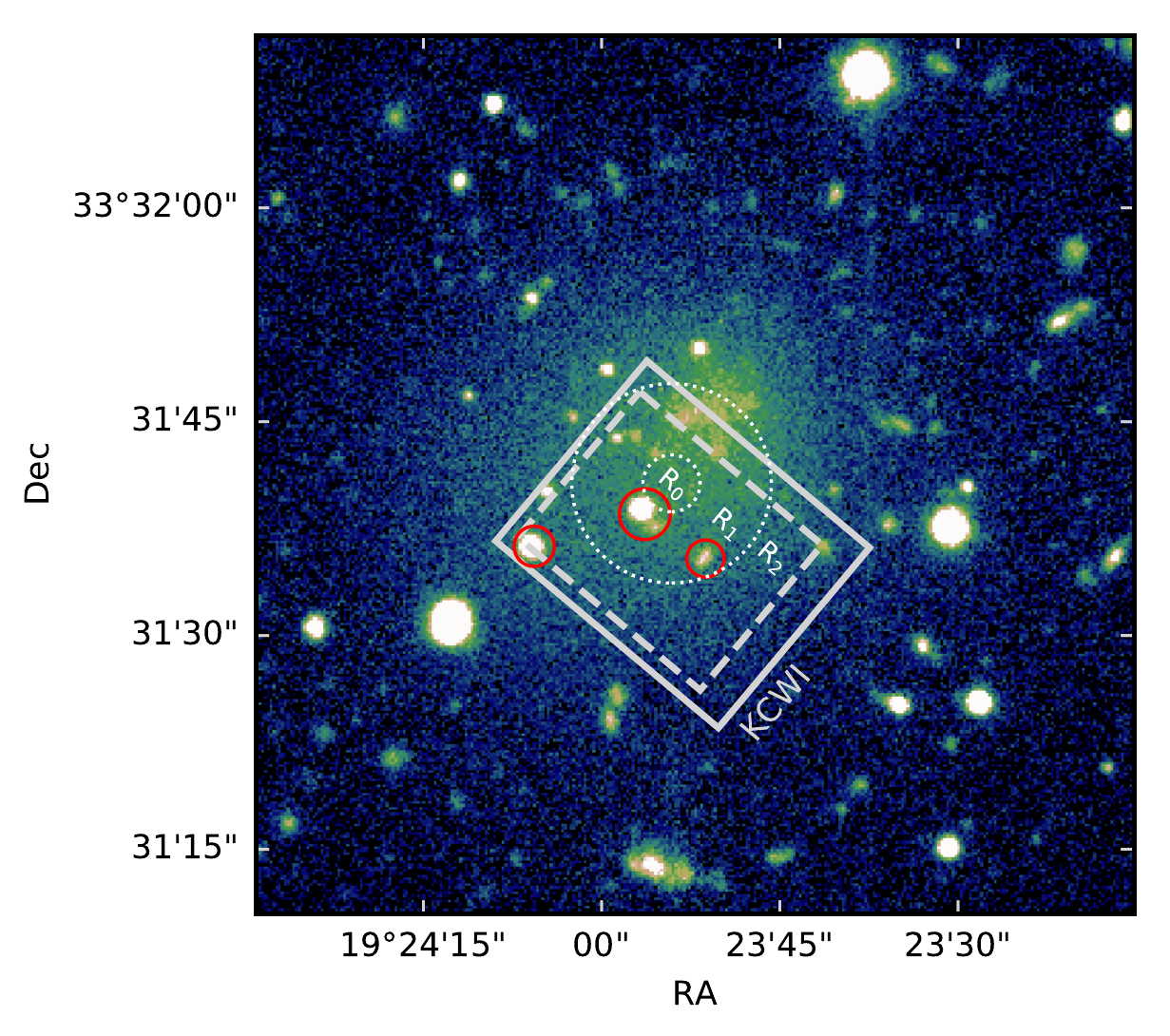}
\end{center}
\caption{Radial apertures and masks. As in Fig.~\ref{fig:spec} but showing the effective field of view of KCWI (dashed line) and the masked background/foreground objects (red circles). The three radial annuli apertures described above (R$_0$, R$_1$, and R$_2$) are indicated with doted white circles.}
\label{fig:regions}
\end{figure}

For clarity, Fig.~\ref{fig:regions} shows the three circular apertures mentioned above. Note that they are completely independent, i.e., they are not {\it cumulative} apertures. We also show the effective field of view after trimming the edges. The masked regions containing visible background/foreground objects are marked in red.

\subsection{Spectrophotometric calibration} \label{sec:fluxcal}

A precise calibration of the KCWI data across all wavelengths is needed in order to robustly measure abundance pattern variations. Relative differences in the flux level could be artificially interpreted as changes in the measured equivalent widths, becoming a potential source of systematic errors in our analysis. In practice, we tried to minimize this effect by combining line-strength over a wide range of wavelengths. In Fig.~\ref{fig:indices} we show how all magnesium lines predict higher metallicities than iron lines. This would be hard to explain solely by flux calibration issues since magnesium-sensitive and iron-sensitive features are spread across the optical range. In order to mimic this effect, any flux calibration issue would have to precisely follow the location of the different lines, changing the relative flux response in such a way that all magnesium and iron lines behave in the same way. Moreover, the Mg\,4780 is located $\sim$200\AA \ away from the other lines, but it shows similar variations. Note also that while measuring the kinematics a multiplicative polynomial was applied to the data in order to match the (flux-calibrated) set of stellar population models. Before measuring any equivalent width, we corrected our data for this additional term. This second-order correction minimizes the effect of systematics introduced by, e.g., the spectral type of our standard star, as stellar population models have been accurately flux calibrated. Finally, it is worth noting that the central band-pass of the Mgb and Mg2 features is roughly the same, but the blue and red pseudo-continua, are significantly different  \citep{Faber85}. We intentionally included these two line-strengths in our analysis to account for flux calibration issues. Since both line indices, probing the same spectral feature but based on different continua show consistent metallicity measurements, our results seem to be robust again uncertainties in the spectrophotometry calibration.

The robustness of our analysis is also evident from Fig.~\ref{fig:empi}. The residuals in this figure empirically show the peculiar [Mg/Fe] abundance ratio of \dgs \ after subtracting the best-fitting model from STECKMAP. This best-fitting model includes a continuum term that corrects for flux calibration issues, which in practice means that the differences shown in Fig.~\ref{fig:empi} are not likely a consequence of the data reduction/calibration.

\begin{figure}
\begin{center}
\includegraphics[width=8.5cm]{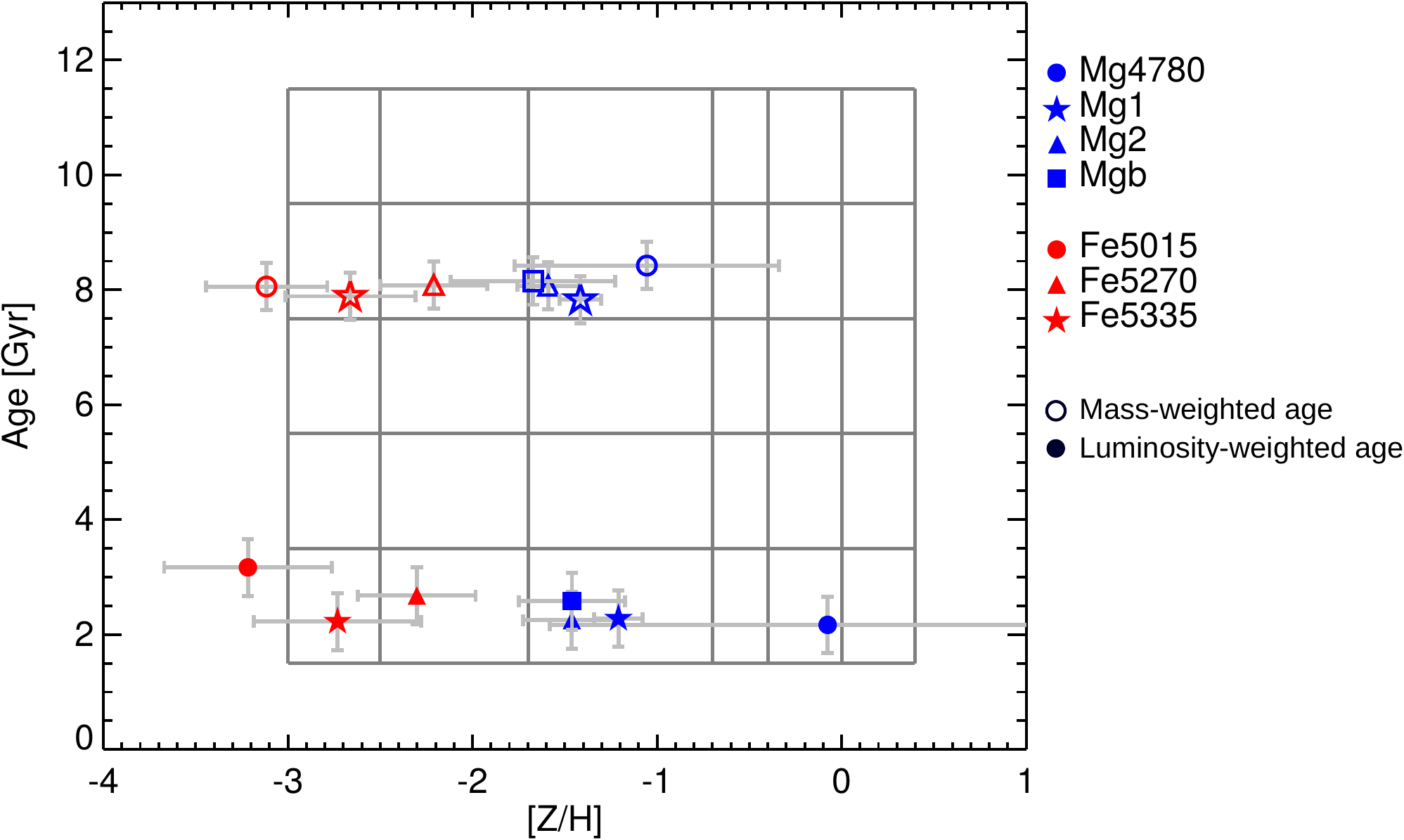}
\end{center}
\caption{Robustness of the [Mg/Fe] measurement: flux calibration. Same as in Fig.~\ref{fig:indices} but with equivalent widths measured after removing the continuum in data and models using the index pseudo-continua definitions. Removing any short-scale variation in the spectrum of \dgs \ does not change our main result, showing that the extreme [Mg/Fe] measurement is not related to systematic uncertainties in the spectrophotometric calibration. Error bars correspond to 1-$\sigma$ uncertainties.}
\label{fig:flux_index}
\end{figure}

Given the sensitivity of our analysis to potential flux calibration issues, we have done three additional tests to complement all the previous arguments. First, we have repeated the analysis shown in  Fig.~\ref{fig:indices}, but this time removing any information encoded in the continuum. The process was the following: for each line-strength index, we used blue and red pseudo-continua to fit a straight line. Then, we divided the spectrum around the index by this best-fitting straight line in order to remove the continuum, and we finally measured the equivalent width. While similar to the standard way of measuring line-strengths, this approach remove any short-scale continuum variation, making the analysis virtually insensitive to flux calibration issues. We did the same (independently) to the PEGASE-HR stellar population templates, and we then compared data and models. Finally, we have also increased and decreased the order of the polynomial used in the kinematics analysis (see above). By doing this we can assess the effect of this correction on the measured abundance pattern. The result of all these tests are presented in Fig~\ref{fig:flux_index}, showing the robustness of our approach. It is worth mentioning that Z$_\mathrm{Mg}$-Z$_\mathrm{Fe}=0$ in Fig~\ref{fig:flux_index} already corresponds to [Mg/Fe] = +0.4, i.e., that of inherited from the (metal-poor) stellar population models. Hence, the measured positive values indicate an even higher [Mg/Fe] value for \dgs.

The separation between magnesium and iron lines changes but not significantly enough to alter our conclusions, further confirming that the extreme [Mg/Fe] value measured for \dgs \ is not driven by a poor flux calibration. The rightmost symbol in Fig.~\ref{fig:tests} shows the difference between magnesium-based and iron-based metallicities measured after removing the index continua, compared to all the other tests described above.

\bsp	
\label{lastpage}
\end{document}